\documentclass[showpacs,aps,graphicx,twocolumn]{revtex4}
\usepackage{amssymb}
\usepackage{txfonts}
\usepackage{graphicx}

\begin{document}

\title{Optimal nonlocal multipartite entanglement concentration  based on projection measurements\footnote{Published in Phys. Rev. A \textbf{85}, 022311 (2012)}}

\author{Fu-Guo Deng\footnote{Email address:
fgdeng@bnu.edu.cn} }
\address{ Department of Physics, Applied Optics Beijing Area Major Laboratory, Beijing Normal University, Beijing 100875, China }
\date{\today }

\date{\today }

\begin{abstract}
We propose an optimal nonlocal entanglement concentration protocol
(ECP) for multi-photon systems in a partially entangled pure state,
resorting to the projection measurement on an additional photon. One
party in quantum communication first performs a parity-check
measurement on her photon in an $N$-photon system and an additional
photon, and then she projects the additional photon into an
orthogonal Hilbert space for dividing the original $N$-photon
systems into two groups. In the first group, the $N$ parties will
obtain a subset of $N$-photon systems in a maximally entangled
state. In the second group, they will obtain some less-entangled
$N$-photon systems which are the resource for the entanglement
concentration in the next round. By iterating the entanglement
concentration process several times, the present ECP has the maximal
success probability which is just equivalent to the entanglement of
the partially entangled state. That is, this ECP is an optimal one.
\end{abstract}
 \maketitle

\section{introduction}

Entanglement plays an important role in quantum information and
quantum computation \cite{book}. For example, the powerful speedup
of quantum computation resorts to multipartite entanglement
\cite{book}. In quantum communication, the two legitimate users, say
the sender Alice and the receiver Bob, can use entangled quantum
systems to transmit a private key \cite{Ekert91,BBM92}. Moreover,
quantum dense coding \cite{densecoding,super2} and quantum
teleportation \cite{teleportation} need entangled quantum systems to
setup the quantum channel. In a long-distance quantum communication,
quantum repeaters are required  because  quantum signals can only be
transmitted over an optical fiber or a free space not more than
several hundred kilometers with current technology, although there
are some quantum key distribution protocols (QKDs) based on single
photons \cite{bb84} or weak pulses
\cite{faint1,faint2,faint3,faint4}. In a practical transmission or
the process for storing an entangled quantum system, it inevitably
suffers from channel noise and its environment. The noise will make
the system decoherent, which will decrease the security of QKD
protocols and the fidelity of quantum teleportation and dense
coding.

Entanglement purification is used to extract some high-fidelity
entangled systems from a less-entangled ensemble in a mixed state
and it has been widely studied
\cite{Bennett1,Deutsch,Pan1,Simon,shengpra,shengpratwostep,shengpraonestep,wangcpra,wangcqic,dengEMEPP}.
since Bennett \emph{et al}. \cite{Bennett1} proposed the original
entanglement purification protocol (EPP) to purify two-photon
systems in a Werner state in 1996. For example, Deutsh \emph{et al}.
\cite{Deutsch} optimized the first EPP with  two additional specific
unitary operations. In 2001, an EPP based on linear optical elements
was introduced by Pan \emph{et al.}\cite{Pan1}. In 2002, Simon and
Pan \cite{Simon} proposed an EPP with a currently available
parametric down-conversion (PDC) source. In 2008, an efficient EPP
\cite{shengpra} based on a PDC source was introduced with cross-Kerr
nonlinearity. In 2010,   a two-step deterministic EPP (DEPP) was
presented. Subsequently, a one-step DEPP \cite{shengpraonestep} was
proposed, only resorting to the spatial entanglement or the
frequency entanglement and linear optical elements. In 2011, Wang
\emph{et al.} proposed an interesting EPP for electron spins of
quantum dots, resorting to microwave cavity \cite{wangcpra} and
another EPP for two-photon systems with cross-Kerr nonlinearity
\cite{wangcqic}. We proposed an efficient multipartite EPP with
cross-Kerr nonlinearity in which the cross-combination items can be
used to distill some entangled subsystems \cite{dengEMEPP}.

Compared with EPPs, entanglement concentration is more efficient for
the two remote parties in quantum communication, say Alice and Bob,
to distill some maximally entangled systems from an ensemble in a
less-entangled pure state because EPPs should consume a great deal
of quantum resource as it can only improve the fidelity of systems
in a mixed entangled state, not obtain a maximally entangled state
directly. Up to now, there are some interesting entanglement
concentration protocols
\cite{Bennett2,Yamamoto,zhao1,shengpraecp,shengsingle,swapping1,swapping2}.
For example, Bennett \emph{et al.} \cite{Bennett2} proposed the
first entanglement concentration protocol (ECP) in 1996 and called
it the Schmidt projection method. In 1999, Bose \emph{et al.}
\cite{swapping1} proposed another  ECP based on entanglement
swapping. Subsequently,  Shi \emph{et al.}\cite{swapping2} presented
a different ECP based on entanglement swapping and a collective
unitary evolution. In 2001, Yamamoto \emph{et al.} \cite{Yamamoto}
and Zhao \emph{et al.} \cite{zhao1} proposed an ECP based on
polarizing beam splitters (PBSs) independently. Also, they completed
its experimental demonstration \cite{zhao2,Yamamoto2}. In 2008, we
proposed an ECP \cite{shengpraecp} by exploiting cross-Kerr
nonlinearities to distinguish the parity of two polarization
photons, resorting to the  Schmidt projection method. By iteration
of the entanglement concentration process, it has a far higher
efficiency and yield than those with linear optical elements
\cite{Yamamoto,zhao1}. In 2010, the first single-photon ECP
\cite{shengsingle} was discussed with cross-Kerr nonlinearity.

All the existing ECPs
\cite{Bennett2,swapping1,swapping2,Yamamoto,zhao1,shengpraecp,shengsingle}
can be divided into two groups. In the first group, the parameters
of the less-entangled pure state $\alpha \vert H\rangle_{A}\vert
H\rangle_{B} + \beta \vert V\rangle_{A}\vert V\rangle_{B}$ are
unknown, such as those in
Refs.\cite{Bennett2,Yamamoto,zhao1,shengpraecp,shengsingle}. In the
other group, the parameters $\alpha$ and $\beta$ are known to Alice
and Bob \cite{swapping1,swapping2}. Here $|H\rangle$ and $|V\rangle$
represent the horizontal and the vertical polarizations of photons.
The subscripts $A$ and $B$ represent the photons hold by Alice and
Bob, respectively. In a practical quantum communication, it is not
difficult for Alice and Bob to obtain  information about the
parameters $\alpha$ and $\beta$ if they measure an enough number of
sample photon pairs. From the view of efficiency, the ECPs based on
a collective unitary evolution \cite{swapping1,swapping2}  are
efficient as their success probability equals to a half of the
entanglement of the less-entangled pure state $E=min\{2|\alpha|^2,
2|\beta|^2\}$, higher than others
\cite{Bennett2,Yamamoto,zhao1,shengpraecp,shengsingle}. However, the
collective unitary evolution is usually difficult to implement in
the experiment and there are no experimental proposals. Moreover,
all existing ECPs
\cite{Bennett2,Yamamoto,zhao1,shengpraecp,shengsingle,swapping1,swapping2}
are, in essence, based on the Schmidt projection method
\cite{Bennett2} and they exploit a pair of multi-qubit partially
entangled systems to obtain a maximally entangled system with the
success probability limit $E$.

In this paper, we proposed an optimal nonlocal ECP for $N$-photon
systems in a known partially entangled pure state, resorting to the
projection measurement on an additional photon. It does not depend
on a pair of systems in a partially entangled state in each round of
concentration, just each system itself and some additional single
photons, which makes it far different from others
\cite{Bennett2,Yamamoto,zhao1,shengpraecp,shengsingle,swapping1,swapping2}.
In the present ECP, one of the parties in quantum communication, say
Alice first performs a parity-check measurement on her photon $A$
and an additional photon $a$, and then she projects the additional
photon into an orthogonal Hilbert space $\{\vert\varphi\rangle,
\vert\varphi^\bot\rangle\}$ for dividing the original $N$-photon
systems into two groups. In the first group, the $N$ parties  in
quantum communication will obtain the $N$-photon systems in a
maximally entangled state when the additional photon is projected
into the state $\vert\varphi^\bot\rangle$. In the second group, they
will obtain some $N$-photon systems in another partially entangled
state, which are the resource for entanglement concentration in the
next round. By iterating the process several times (usually no more
than three times), the present ECP has a success probability $P$
which is nearly equivalent to the entanglement of the partially
entangled state $E$, twice of those based entanglement swapping and
a collective unitary evolution \cite{swapping1,swapping2}. Moreover,
it does not require a collective unitary evolution, which decreases
the difficulty of its implementation.

\section{optimal nonlocal multipartite entanglement concentration  based on projection measurements}

Our ECP is based on a parity-check detector (PCD) and the projection
measurement on an additional photon. We first introduce the
principle of the PCD based on cross-Kerr nonlinearity below and then
our ECP for two-photon systems. In fact, the PCD here is similar to
those in Refs.\cite{QND1,dengEMEPP}.

\begin{figure}[!h]
\begin{center}
\includegraphics[width=6cm,angle=0]{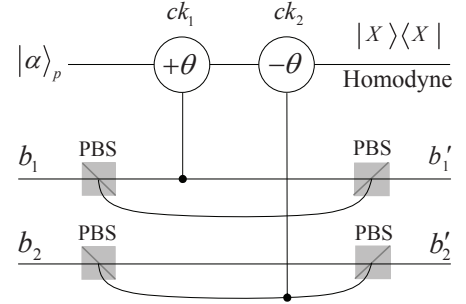}
\caption{The principle of a parity-check detector (PCD), the same as
that in Ref.\cite{dengEMEPP}.  PBS  represents a polarizing beam
splitter which transmits horizontal polarization $\vert H\rangle$
 and reflects the vertical polarization $\vert V \rangle$. $\pm
\theta$ represent two cross-Kerr nonlinear media which introduce the
phase shifts $\pm \theta$ when there is a photon passing through the
media. $|X\rangle\langle X|$ represents an X quadrature
measurement.} \label{fig1_QND}
\end{center}
\end{figure}

The Hamiltonian of a cross-Kerr nonlinearity is \cite{QND1}
\begin{eqnarray}
H_{ck}=\hbar\chi a^{+}_{s}a_{s}a^{+}_{p}a_{p}.
\end{eqnarray}
Here $a^{+}_{s}$ and $a^{+}_{p}$ are the creation operations, and
$a_{s}$ and $a_{p}$ are the destruction operations. $\chi$ is the
coupling strength of the nonlinearity. If a signal state
$|\Psi\rangle_s=c_{0}|0\rangle_{s}+c_{1}|1\rangle_{s}$
($|0\rangle_{s}$ and $|1\rangle_{s}$ denote that there are no photon
and one photon respectively in this state) and a coherent probe beam
in the state $|\alpha\rangle_p$ couple with a cross-Kerr
nonlinearity medium, the evolution of the whole system  can be
described as:
\begin{eqnarray}
U_{ck}|\Psi\rangle_{s}|\alpha\rangle_{p}&=&
e^{iH_{ck}t/\hbar}[c_{0}|0\rangle_{s}+c_{1}
|1\rangle_{s}]|\alpha\rangle_{p} \nonumber\\
&=& c_{0}|0\rangle_{s}|\alpha\rangle_{p}+c_{1}|1\rangle_{s}| \alpha
e^{i\theta}\rangle_{p},
\end{eqnarray}
where $\theta=\chi t$ and $t$ is the interaction time. The coherent
beam picks up a phase shift $\theta$ directly proportional to the
number of the photons in the Fock state $|\Psi\rangle_s$. Based on
this feature of a cross-Kerr nonlinearity, the principle of our PCD
is shown in Fig.\ref{fig1_QND}. With an X quadrature measurement in
which the the states $\vert \alpha e^{\pm i\theta} \rangle_p$ cannot
be distinguished \cite{QND1,QND3}, one can distinguish
superpositions and mixtures of $|HH\rangle$ and $|VV\rangle$ from
$|HV\rangle$ and $|VH\rangle$ as the probe beam $\vert \alpha
\rangle_p$ will pick up a phase shift $\theta$ if the two photons is
in the state $|HH\rangle_{b_1b_2}$ or $|VV\rangle_{b_1b_2}$. If it
picks up a phase shift $0$, the two photons are in the state
$|VH\rangle_{b_1b_2}$ or $|HV\rangle_{b_1b_2}$. That is, when the
parity of the two photons is odd, the coherent beam will pick up a
phase shift $0$; otherwise it will pick up  a phase shift $\theta$.

\begin{figure}[!h]
\begin{center}
\includegraphics[width=8cm,angle=0]{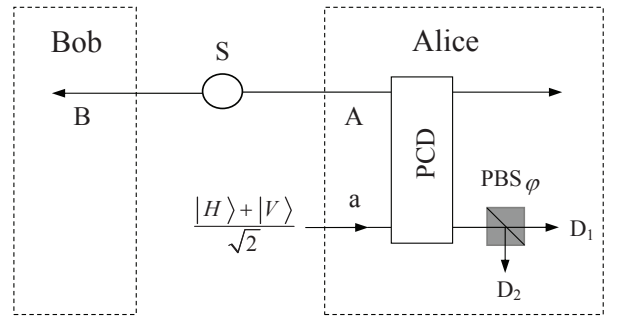}
\caption{The schematic diagram of the present entanglement
concentration protocol for two-photon systems in a less-entangled
pure state. The pair of identical less-entanglement photons $A_1$
and $B_1$ are sent to Alice and Bob from source ($S$), respectively.
PCD represents a parity-check detector. }\label{fig2}
\end{center}
\end{figure}

With the PCD shown in Fig.\ref{fig1_QND},  the principle of our ECP
for two-photon systems in a less-entangled pure state  is shown in
Fig.\ref{fig2}.  Suppose the photon pair $AB$ is initially in the
following polarization less-entangled pure state:
\begin{eqnarray}
|\Phi_1\rangle_{AB}=\alpha|H\rangle_{A}|H\rangle_{B} +
\beta|V\rangle_{A}|V\rangle_{B},\label{originalstate1}
\end{eqnarray}
where $\alpha$ and $\beta$ are two real numbers and $
|\alpha|^{2}+|\beta|^{2}=1$. The same as the ECPs with entanglement
swapping \cite{swapping1,swapping2}, Alice and Bob know these two
parameters before they distill a subset of maximally entangled
photon pairs from a set of photon pairs in the state
$|\Phi_1\rangle_{AB}$.

For distilling some maximally entangled photon pairs, Alice prepares
an additional photon $a$ in the polarization state $\vert
\Phi\rangle_{a}=\frac{1}{\sqrt{2}}(\vert H\rangle + \vert V\rangle)$
and then performs a parity-check measurement on her photons $A$ and
$a$. If she obtains an even parity, the three-photon system  $ABa$
is in the state
\begin{eqnarray}
\vert\Psi_e\rangle_{ABa}=\alpha|H\rangle_{A}|H\rangle_{B}\vert
H\rangle_{a} + \beta|V\rangle_{A}|V\rangle_{B}\vert V\rangle_{a}.
\end{eqnarray}
If she obtains an odd parity, the system is in the state
\begin{eqnarray}
\vert\Psi_o\rangle_{ABa}=\alpha|H\rangle_{A}|H\rangle_{B}\vert
V\rangle_{a} + \beta|V\rangle_{A}|V\rangle_{B}\vert H\rangle_{a},
\end{eqnarray}
and Alice can transform it into the state $\vert\Psi_e\rangle_{ABa}$
by performing a bit-flip operation $\sigma_x=|H\rangle\langle V| +
|V\rangle\langle H|$ on the photon $a$. That is, we need only
describe the principle of the present ECP when Alice and Bob obtain
their photon systems in the state $\vert\Psi_e\rangle_{ABa}$ below.

We can rewrite the state $\vert\Psi_e\rangle_{ABa}$ under the
orthogonal basis $\{ \vert \varphi_1\rangle_{a}=\alpha \vert
H\rangle - \beta \vert V\rangle,  \vert
\varphi^\bot_1\rangle_{a}=\beta \vert H\rangle + \alpha\vert
V\rangle\}$, that is,
\begin{eqnarray}
\vert\Psi_e\rangle_{ABa} &=& (\alpha^2\vert H\rangle_A|H\rangle_B -
\beta^2\vert
V\rangle_A|V\rangle_B)\vert \varphi_1\rangle_a \nonumber\\
&+&  \sqrt{2}\alpha\beta \cdot \frac{\vert H\rangle_A|H\rangle_B +
\vert V\rangle_A|V\rangle_B}{\sqrt{2}}\vert
\varphi^\bot_1\rangle_{a}.
\end{eqnarray}
Alice can use a PBS$_\varphi$, whose optical axis is placed at the
angle $\varphi_1$, and two detectors to complete the measurement on
the additional photon $a$ with the basis $\{\vert
\varphi_1\rangle_{a}, \vert \varphi^\bot_1\rangle_{a}\}$, shown in
Fig.\ref{fig2}. Here $cos\varphi_1=\alpha$ and
$sin\varphi_1=-\beta$. If Alice obtains the state $\vert
\varphi^\bot_1\rangle_{a}$ when she measures the additional photon
$a$, the photon pair $AB$ is in the maximally entangled state
$\vert\phi^+\rangle_{AB}=\frac{1}{\sqrt{2}}(\vert HH\rangle + \vert
VV\rangle)_{AB}$, which takes place with the probability of
$2\alpha^2\beta^2$. If Alice obtains the state $\vert
\varphi_1\rangle_{a}$, the photon pair $AB$ is in another partially
entangled pure state (without normalization)
\begin{eqnarray}
\vert\Phi_2\rangle_{AB} &=& \alpha^2\vert H\rangle_A|H\rangle_B -
\beta^2\vert V\rangle_A|V\rangle_B,
\end{eqnarray}
which takes place with the probability of $\alpha^4 +
\beta^4=1-2\alpha^2\beta^2$.

\begin{figure}[!h]
\begin{center}
\includegraphics[width=8cm,angle=0]{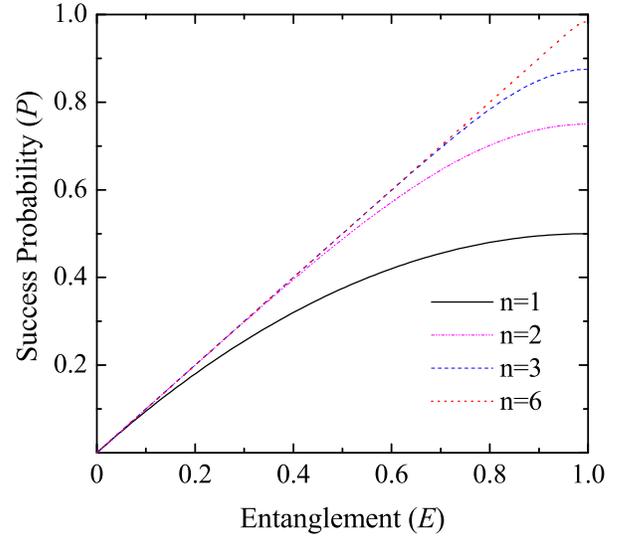}
\caption{(Color online) The relation between the success probability
of the present ECP $P$ and the entanglement of the partially
entangled state $E$ under the iteration numbers of entanglement
concentration $n=$ 1, 2, 3, and 6, respectively. }\label{fig3}
\end{center}
\end{figure}

It is obvious that the less-entangled pure state
$\vert\Phi_2\rangle_{AB}$ has the same form as the state
$\vert\Phi_1\rangle_{AB}$ shown in Eq.(\ref{originalstate1}). We
need only replace $\alpha$ and $\beta$ with $\alpha'\equiv
\frac{\alpha^2}{\sqrt{\alpha^4 +  \beta^4}}$ and $\beta' \equiv -
\frac{\beta^2}{\sqrt{\alpha^4 +  \beta^4}}$, respectively. That is,
Alice and Bob can distill the maximally entangled state
$\vert\phi^+\rangle_{AB}$ from the state $\vert\Phi_2\rangle_{AB}$
with the probability of $2(\alpha^4 + \beta^4)\alpha'^2\beta'^2$ by
adding another additional photon $a_1$ and a parity-check
measurement. Moreover, they can distill the photon pairs in the
maximally entangled state $\vert\phi^+\rangle_{AB}$ from the
less-entangled systems in the next round yet. That is, by iterating
the entanglement concentration process $n$ times, the total success
probability of this ECP is
\begin{eqnarray}
P_n &=& 2[\frac{\alpha^2\beta^2}{\alpha^2 + \beta^2} +
\frac{\alpha^4\beta^4}{(\alpha^2 + \beta^2)(\alpha^4 + \beta^4)}
\nonumber\\
&+& \frac{\alpha^8\beta^8}{(\alpha^2 + \beta^2)(\alpha^4 +
\beta^4)(\alpha^8 + \beta^8)} \nonumber\\
&+& \frac{\alpha^{16}\beta^{16}}{(\alpha^2 + \beta^2)(\alpha^4 +
\beta^4)(\alpha^8 + \beta^8)(\alpha^{16}+\beta^{16})} +
\cdots\nonumber\\
&+& \frac{\alpha^{2^n}\beta^{2^n}}{(\alpha^2 + \beta^2)(\alpha^4 +
\beta^4)(\alpha^8 + \beta^8)\cdots(\alpha^{2^n}+\beta^{2^n})}].
\label{totalprobability}
\end{eqnarray}

If $|\alpha| \leq |\beta|$, the entanglement of the  state
$|\Phi_1\rangle_{AB}=\alpha|H\rangle_{A}|H\rangle_{B} +
\beta|V\rangle_{A}|V\rangle_{B}$ is $E=2|\alpha|^2$
\cite{swapping1,swapping2}. The relation between the success
probability $P$ that the two parties obtain a photon pair $AB$ in
the maximally entangled state $\vert \phi^+\rangle_{AB}$ from a pair
in the partially entangled state $|\Phi_1\rangle_{AB}$ and the
entanglement $E$ is shown in Fig.\ref{fig3}. When $E<0.4$, Alice and
Bob need only perform the entanglement concentration process twice
($n=2$) for obtaining the success probability $P$ nearly equivalent
to the entanglement $E$. When $0.4<E<0.72$, they should perform the
process three times for obtaining an optimal success probability.
From Fig.\ref{fig3}, one can see that  six times for the iteration
of the entanglement concentration process is enough to obtain an
optimal success probability.

\section{Discussion and summary}

It is straightforward to generalize our ECP to reconstruct maximally
entangled $N$-photon GHZ states from partially entangled GHZ-class
states.  Suppose the partially entangled $N$-photon GHZ-class states
are described as follows:
\begin{eqnarray}
|\Phi_N\rangle=\alpha|HH\cdot\cdot\cdot H\rangle_{AB\dots Z} +
\beta|VV\cdot\cdot\cdot V\rangle_{AB\dots Z},
\end{eqnarray}
where $|\alpha|^{2}+|\beta|^{2}=1$. The subscript $A$, $B$, $\dots$,
and $Z$ represent the photons held by Alice, Bob, $\dots$, and Zach,
respectively. If we define $\vert H'\rangle_{N'} \equiv
|H\cdot\cdot\cdot H\rangle_{B\dots Z}$ and $\vert V'\rangle_{N'}
\equiv |V\cdot\cdot\cdot V\rangle_{B\dots Z}$, the state
$|\Phi_N\rangle$ can be rewritten as
\begin{eqnarray}
|\Phi_N\rangle=\alpha|H\rangle_{A}\vert H'\rangle_{N'} +
\beta|V\rangle_{A}\vert V'\rangle_{N'}.
\end{eqnarray}
It has the same form as the state $\vert \Phi_1\rangle_{AB}$ shown
in Eq.(\ref{originalstate1}). So the $N$ parties can also obtain the
maximally entangled $N$-photon systems with the total success
probability $P$ if Alice deals with the photon $A$ in the $N$-photon
system and another additional photon $a_2$ in the same way as the
case with nonmaximally entangled two-photon pure state.

\begin{figure}[!h]
\begin{center}
\includegraphics[width=8cm,angle=0]{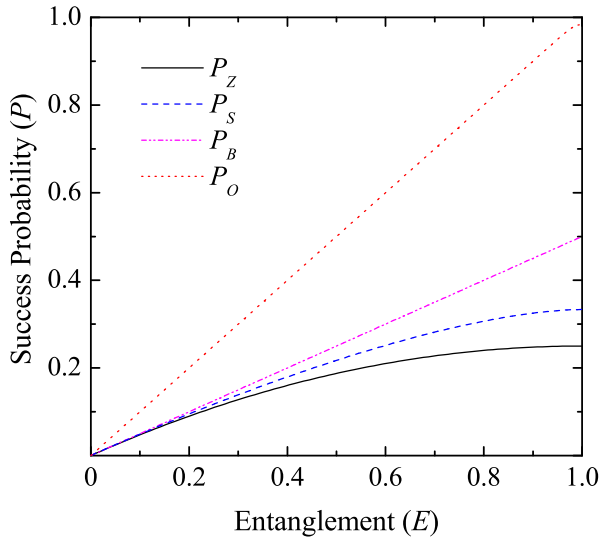}
\caption{(Color online) The comparison of the success probability
between  the present ECP  and other ECPs. $P_O$ is the success
probability in the present ECP.  $P_Z$,  $P_S$, and $P_B$ are the
success probabilities for each system in the works by Zhao \emph{et
al.} \cite{zhao1} and Yamamoto \emph{et al.} \cite{Yamamoto}, Sheng
\emph{et al.} \cite{shengpraecp}, and Bose \emph{et al.}
\cite{swapping1} and Shi \emph{et al.} \cite{swapping2},
respectively.}\label{fig4}
\end{center}
\end{figure}

The comparison of the success probabilities between the present ECP
and other ECPs
\cite{Bennett2,swapping1,swapping2,Yamamoto,zhao1,shengpraecp,shengsingle}
is shown in Fig.\ref{fig4}. Here $P_Z$ represents the success
probability in the ECP by Zhao \emph{et al.} \cite{zhao1} and
Yamamoto \emph{et al.} \cite{Yamamoto}. It is just the one in the
ECP based on the Schmidt projection method by Bennett \emph{et al.}
\cite{Bennett2}. $P_O$ and $P_S$ represent the success probability
in the present ECP and that in the ECP with cross-Kerr nonlinearity
\cite{shengpraecp} under the iteration number $n=6$, respectively.
$P_B$ is the success probability in the ECPs by Bose \emph{et al.}
\cite{swapping1} and Shi \emph{et al.} \cite{swapping2} in an ideal
condition. All other existing ECPs
\cite{Bennett2,Yamamoto,zhao1,shengpraecp,shengsingle,swapping1,swapping2}
are based on the Schmidt projection method although some exploit a
collective unitary  evolution to improve their success probability,
which requires at least a pair of systems in a partially entangled
state to distill a system in a maximally entangled state with some
probability. For each system in a partially entangled original
state, the maximal success probability is just a half of the
entanglement of the initial state $E$. However, the present ECP does
not depend on a pair of systems in a partially entangled state in
each round of concentration, just each system itself and some
additional single photons, which makes it far different from others
\cite{Bennett2,Yamamoto,zhao1,shengpraecp,shengsingle,swapping1,swapping2}.
Moreover, it is obvious that the present ECP is far more efficient
than others as its success probability equals to the entanglement of
the partially entangled state. That is, the present ECP is an
optimal one.

The present ECP requires that the parties obtain the information
about the initial state, as the same as those in
Refs.\cite{swapping1,swapping2}, but different from those in
Refs.\cite{Bennett2,Yamamoto,zhao1,shengpraecp}. On one hand, those
ECPs \cite{Bennett2,Yamamoto,zhao1,shengpraecp}, which do  not
require the parties know accurately the information about the
initial state, can be used to concentrate nonlocally the systems in
a partially entangled known state. Especially, the ECP based on
linear optical elements \cite{Yamamoto,zhao1} and the efficient ECP
based on nonlinear optics \cite{shengpraecp} give a detailed way for
its implementation, which are different from the ones based on a
collective unitary evolution. On the other hand, all the existing
ECPs
\cite{Bennett2,Yamamoto,zhao1,shengpraecp,shengsingle,swapping1,swapping2}
are, in essence, based on the Schmidt projection method in which a
two-photon system in the state $\alpha\vert H\rangle_{A_1}\vert
H\rangle_{B_1} + \beta\vert V\rangle_{A_1}\vert V\rangle_{B_1}$ and
another one in the state  $\beta\vert H\rangle_{A_2}\vert
H\rangle_{B_2} + \alpha\vert V\rangle_{A_2}\vert V\rangle_{B_2}$ are
used to distill a two-photon system in the state
$\frac{1}{\sqrt{2}}(\vert H\rangle\vert H\rangle + \vert
V\rangle\vert V\rangle)_{A_1B_2}$ with an average success
probability of $|\alpha\beta|^2$ or $|\alpha|^2$ (if $|\alpha|\leq
|\beta|$) for each system.

The key element in the present ECP is the PCD. We construct the PCD
with cross-Kerr nonlinearity. At present, the implementation of a
clean cross-Kerr nonlinearity is still  difficult in the experiment,
especially  with natural cross-Kerr nonlinearities. Fortunately, the
PCD in our ECP does not require a large nonlinearity and it works
for small values of the cross-Kerr coupling, which decreases the
difficulty of its implementation  \cite{QND1,QND3}. On the one hand,
the fidelity of the PCD with cross-Kerr nonlinearity in a practical
application at present can not be improved to be a unit as there are
always phase noises \cite{ck1,ck2,ck3}. On the other hand, a great
number of works are focused on the photon-photon nonlinear
interaction \cite{He,friedler1}, which provides many ways for
constructing the PCD, such as these based on quantum dot spins in
micro-wave cavity \cite{dot1,dot2}, a Rydberg atom ensemble
\cite{Rydberg}, a cavity waveguide \cite{cavity_waveguide},
hollow-core waveguides \cite{hollow_core_waveguide}, and so on. We
use the PCD based on  cross-Kerr nonlinearity to describe the
principle of our ECP. It works with the PCDs based on other
nonlinear interactions.

In summary, we have proposed  an optimal nonlocal ECP for
multipartite  partially entangledstates, resorting to projection
measurements. Alice exploits the PCD based on cross-Kerr
nonlinearity to extend the partially entangled $N$-photon system to
an $(N+1)$-photon system first and then she projects the additional
photon with a suitable orthogonal basis. By detecting the state of
the additional photon, the $N$-parties in quantum communication can
divide their $N$-photon systems into two groups. One is in the
maximally entangled state and the other is in another partially
entangled state which is just the resource for the entanglement
concentration in the next round. By iterating the entanglement
concentration process several times, the $N$ parties can obtain a
subset of $N$-photon systems in the maximally entangled state with
the maximal success probability which is just equivalent to the
entanglement of the partially entangled state. Compared with other
ECPs
\cite{Bennett2,swapping1,swapping2,Yamamoto,zhao1,shengpraecp,shengsingle},
the present ECP has the optimal success probability, without
resorting to a collective unitary evolution \cite{swapping2}.

\section*{ACKNOWLEDGMENTS}

This work is supported by the National Natural Science Foundation of
China under Grant No. 10974020 and No. 11174039, NCET, and the
Fundamental Research Funds for the Central Universities.

\end{document}